\documentclass{optica-article}

\journal{opticajournal} 

\articletype{Research Article}

\usepackage{lineno}

\begin{document}

\title{Finesse evolution in a bichromatic cavity under blue and red resonant illumination in vacuum}
\author{M. S. Caracas Nu\~nez,\authormark{1,$\dagger$} A. Kadakkavattathu,\authormark{2,1,3}, G. Gatta\authormark{3,1}, L. Duca, \authormark{3,1} N. Mizukami,\authormark{3,1} and C. Sias\authormark{3,1,4,*}}

\address{\authormark{1}European Laboratory for Nonlinear Spectroscopy (LENS), Via Nello Carrara 1, Sesto Fiorentino, 50019, Italy\\
\authormark{2}Politecnico di Torino, Via Duca degli Abruzzi 24, Turin\\
\authormark{3}Istituto Nazionale di Ricerca Metrologica (INRIM), Viale delle Idee 22, Sesto Fiorentino 50019, Italy\\
\authormark{4}Istituto Nazionale di Ottica CNR-INO, Via Nello Carrara 1, 50019 Sesto Fiorentino 50019, Italy}

\email{\authormark{$\dagger$}michael.caracas@lens.unifi.it}
\email{\authormark{*}c.sias@inrim.it} 


\begin{abstract*} 
The long-term performance of high-finesse optical cavities depends on the stability of their mirror coatings under vacuum and under continuous exposure to high circulating power. In cavities supporting multiple wavelengths, a central question is whether degradation at one wavelength compromises performance at another. Understanding this interplay is essential for those cavity quantum electrodynamics experiments that need diverse optical fields for cavity stabilization, trapping, and light--matter coupling.
Here, we investigate wavelength-dependent degradation and recovery of a bichromatic cavity with high-reflectivity coating at $455$~nm and $671$~nm in vacuum. Cavity ring-down measurements reveal a decrease in the $455$~nm
finesse from $7.7\times10^3$ to $1.9\times10^3$ after $11$~hours of continuous exposure to high circulating power. Simultaneous measurements show no measurable degradation at $671$~nm. Continuous operation at $671$~nm, instead, produces no measurable degradation at either wavelength under the tested conditions. Comparison of oxygen depletion and Avrami-type loss models indicates kinetics compatible with nucleation and growth, without uniquely identifying the microscopic mechanism. We demonstrate partial recovery of the finesse with illumination at 455~nm in air. Our results indicate a wavelength dependence of laser-induced optical damage in bichromatic cavities and provide an estimate of the blue light intensity that can be safely used in such cavities.

\end{abstract*}

\section{Introduction}
High-finesse optical cavities placed under vacuum are critical tools in atomic molecular and optical (AMO) physics. Thanks to their high finesse and narrow cavity linewidth, they can provide the strong light-matter coupling required to interface the narrow optical transitions of single trapped ions, neutral atoms, or solid-state emitters with photons~\cite{Kimble1998}. As such, they play a key role in fundamental physics and metrology, not only as a precise frequency reference~\cite{Kessler2011} but also as tool for metrological enhancements via squeezing in optical atomic clocks~\cite{PedrozoPenafiel2020,Robinson2024} or as a mean to induce long interactions among particles for quantum simulations~\cite{Landig2016, Sauerwein2023}. In quantum technology applications, instead, such systems serve as key building blocks for efficient entangled single-photon sources~\cite{Hijlkema2007,Stute2012} and nodes in optical quantum networks~\cite{Ritter2012, Krutyanskiy2023, Buki2026}. In these systems, bichromatic and multi-wavelength optical cavities - cavities that have a high reflectivity for more than one frequency range - are becoming more and more common with the increasing complexity of AMO experiments and quantum technologies. They are used for various purposes, from providing narrow ULE frequency references of several laser sources~\cite{Barbiero2022}, to decoupling the stabilization of the cavity from the coupling of atoms/ions to the relevant cavity field~\cite{Ye1999, Lee14}.

In all these applications, placing high-reflectivity cavity mirrors inside high-vacuum environments is necessary to mitigate spurious thermal, acoustic, and pressure fluctuations which degrade the cavity finesse and the cavity stability and, in turn, the precision and accuracy of the cavity frequency reference. Despite this clear advantage of vacuum environment, the long-term performance of high-finesse optical cavities under vacuum is frequently degraded. In fact, thin-film dielectric coatings — typically consisting of alternating layers of silica ($\mathrm{SiO_2}$) and tantalum pentoxide ($\mathrm{Ta_2O_5}$) — have been observed to undergo significant degradation over time when placed under vacuum~\cite{Demiryont85,Brandstatter2013,Rudelis:23} and also when illuminated with laser light~\cite{Wagner,Kunz2000, KOSTER2002,Pereira2009}. This is especially common for cavities with high-reflectivity in the ultraviolet (UV) and near UV (NUV) where degradation is usually more severe. Two main causes of degradation are reported in literature. The first concerns vacuum-induced oxygen depletion in the surface layer, identified as a common mechanism driving optical loss in both the ultraviolet and visible/near infrared (NIR) regimes~\cite{Brandstatter2013, Gangloff:15, ballance2017cavity, Rudelis:23}. This effect is purely vacuum related and can be affected by temperature~\cite{Brandstatter2013,Gangloff:15}. In all the works the degradation is reversible with oxygen exposure and can be strongly reduced with the choice of the topmost coating layer~\cite{Burkley:21}. 
The second cause is hydrocarbon contaminants that photochemically react under intense illumination and alter the micro-chemical environment of the coating~\cite{Hollenshead2006, Burkley:21, Brand2023, Das26}. This relates to the combination of vacuum environment with a high intracavity intensity which circulates in the resonator for an extended period of time. The degradation caused by hydrocarbons and its recovery with oxygen has been observed~\cite{Burkley:21, Brand2023, Das26}. Nevertheless, just a few studies report on the finesse degradation after cavity illumination for extended periods of time~\cite{Das26}, mostly because the degradation is extremely fast in the deep UV regime~\cite{Brand2023, Keller}, and this makes it hard to compare the degradation with the hydrocarbon nucleation model. Some studies have already shown the possibility of preventing degradation with clean room facilities~\cite{Schmitz2019}, with micro-chemical environment control of the topmost coating layer~\cite{Brand2023} and or with initial UV conditioning in an oxygen environment~\cite{Burkley:21}.

Although cavity degradation has been documented across several individual spectral regions, systematically testing multi-wavelength cavities, and especially those designed to maintain simultaneously high finesse for a near ultraviolet radiation or blue alongside a secondary visible or near-infrared band, remains an unaddressed challenge. Consequently, it remains unclear whether vacuum and/or light-induced damage localized on the high-reflectivity coating layer translates into optical losses at the secondary wavelength for which the dielectric stack has also high reflectivity and, viceversa, if high power in the VIS/NIR affects the finesse in the shorter wavelength range. Resolving this issue is increasingly vital for novel cutting-edge quantum systems. 

Here, we investigate the degradation of a bichromatic high finesse Fabry-Pérot cavity placed under vacuum that has a high reflectivity at $455$~nm and $671$~nm. The cavity has been developed to address Ba$^+$ ions and Li atoms optical transitions~\cite{Perego2020, Berto}. We observe the long-term effect of a high intracavity circulating power either at $455$~nm or $671$~nm on the finesse measured at both $455$~nm and $671$~nm via cavity ring-down. In this way, we test whether light-induced changes in optical losses affect both high reflectivity bands and find what mechanism governs the degradation of such optical cavity.



\section{Methods}
\subsection{Finesse detection and experimental setup}

When the reflectivity $\mathcal{R}$ of the cavity mirrors is sufficiently high, the cavity round-trip time is much smaller than the decay time of the cavity $\tau$, $2 L/c\ll\tau$ with $L$ being the cavity length and $c$ the speed of light. In this scenario, cavity ring-down measurements provide a direct measure of the total optical losses of a cavity~\cite{Romanini1997,Rempe92}. 
We monitor the decay of the light stored in the cavity by using a photodiode in transmission after the input light is suddenly turned off. The resulting exponential decay of the cavity transmission signal $T(t)$, as detected by the photodiode, is described by
\begin{equation}
    T(t)=A + B\cdot e^{-\frac{(t-t_0)}{\tau}},
    \label{eq:transmission}
\end{equation}
where $A$ is the offset of the transmission signal on the photodiode, $B$ is the initial transmission value, $t_0$ is the time the decay begins. 

For two identical high reflectivity mirrors, the cavity lifetime $\tau$ is linked to the reflectivity $\mathcal{R}$ of the cavity mirrors according to
\begin{equation}
    \tau = -\frac{L}{c\ln\mathcal{R}}
    \simeq \frac{L}{c(1-\mathcal{R})}.
    \label{eq:lifetime}
\end{equation}

The lifetime obtained from a fit to Eq.~\eqref{eq:transmission} therefore allows one to extract the finesse $\mathcal{F}$ of the cavity through
\begin{equation}
    \mathcal{F} = 2\pi\,\mathrm{FSR}\,\tau,
    \qquad
    \mathrm{FSR}=\frac{c}{2L}.
    \label{eq:finesse}
\end{equation}

In our experiment we use a linear high finesse optical cavity with a high reflectivity coating at $455$~nm and $671$~nm. It is composed of two superpolished curved mirrors with radii of curvature $50$~mm with a custom high reflectivity coating (FiveNine Optics) which is composed of layers of $\mathrm{SiO_2}$ and $\mathrm{Ta_2O_5}$, last one being  $\mathrm{Ta_2O_5}$. The nominal losses are $180$~ppm for $455$~nm and $671$~nm  wavelengths. The cavity has a nominal mirror separation of $L \simeq 40$~mm and it is operated in vacuum. For the cavity used in this work, the expected cavity lifetime $\tau$ is on the order of a few microseconds. The cavity is mounted inside a UHV chamber, a turbo pump is used to lower the chamber pressure initially down to about $10^{-6}$~mbar, which then slowly decreases down to $10^{-9}$~mbar over the course of the days. The pressure is monitered by using a gauge.

\begin{figure}[ht]
    \centering
    \includegraphics[width=1\linewidth]{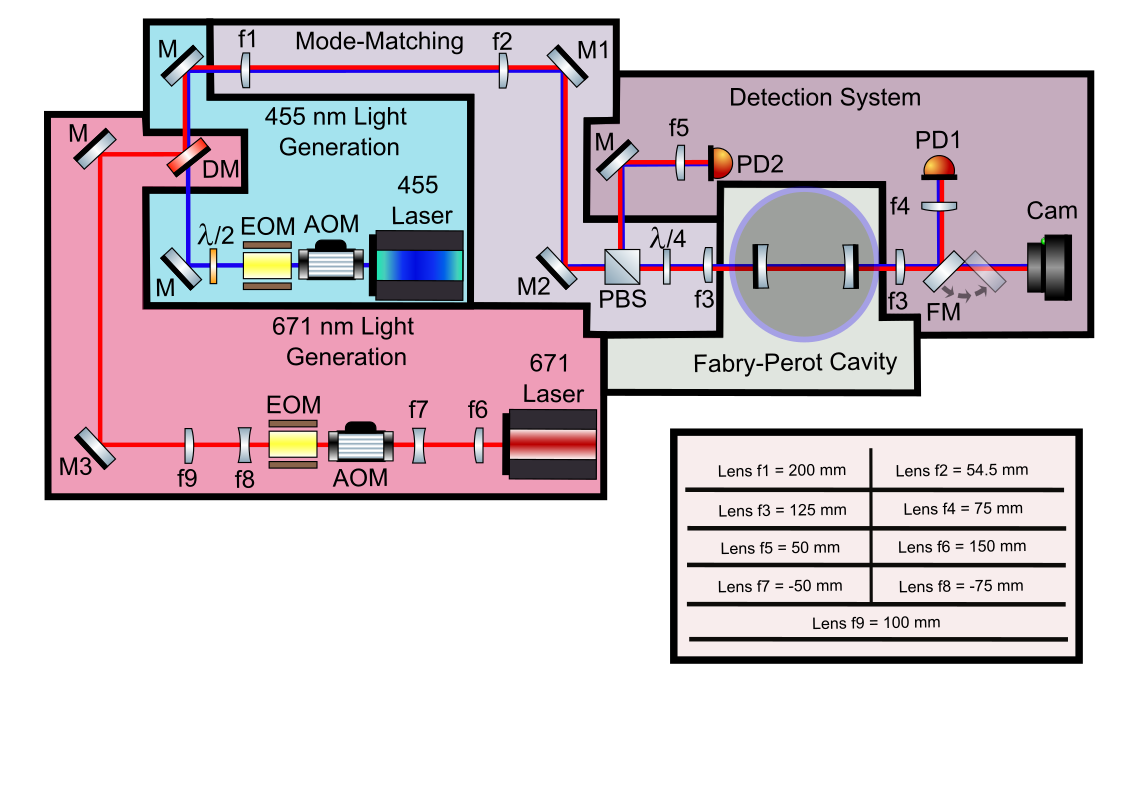}
    \caption{Experimental setup: The $455$~nm light is combined with the $671$~nm on a dichroic mirror (DM). The combined beams are mode-matched to the Fabry–Pérot cavity using lenses $f_1$, $f_2$ and $f_3$ and the dielectric mirrors (M1 and M2). Photodiodes PD1 and PD2 detect the transmitted and reflected cavity signals, respectively, along with a Thorlabs camera CS165MU1/M. A flip mirror (FM) directs the transmitted beam to either the camera or PD1. Acousto-optic modulators (AOM) are used for switching off and on the light coupled to the cavity in order to perform the ring-down measurements. Electro-optic modulators (EOMs) are used to generate the error signals.}
    \label{fig:setup}
\end{figure}

The cavity finesse evolution is tested with two lasers, a home-made external cavity diode laser operating at $455$~nm, and a Toptica external cavity diode laser (DL pro) at $671$~nm. The optical setup used for coupling and locking to the optical cavity is illustrated in Fig.~\ref{fig:setup}. The $455$~nm and $671$~nm beams are combined on a short-pass dichroic mirror (DM) and co-propagate in the optical cavity. Both wavelengths setups include an acousto-optic modulator (AOM), and an electro-optic modulator (EOM) for fast switching and error-signal generation. Mode-matching lenses and incoupling lenses are also shown in figure and allowed us to couple about 35\% of the $455$~nm and $671$~nm light into the fundamental transverse mode (TEM$_{00}$) before starting the degradation measurements. The TEM$_{00}$ mode was identified from the observed transmission via a Thorlabs CMOS camera (CS165MU1/M) and this resonance was then used for locking.
Cavity-reflected light is routed to the photodetector PD2. The $455$~nm or the $671$~nm photodetection signal is then mixed with the RF reference driving the corresponding EOM in order to generate the error signal needed for locking the laser onto the cavity. The error signal is then fed to a Moku PID controller to lock the lasers to the selected cavity resonance. At the cavity output, instead, a flip-mirror assembly (FM) allows us to detect the light either on a transmission photodetector (PD1) for recording the cavity ring-down, or on the camera for monitoring the cavity mode. Thorlabs PDA10A2 photodetectors were used for optical detection.

In order to test the cavity degradation dynamics, we keep the lasers, either at $455$~nm or $671$~nm, locked to the cavity for an extended period of time so that the cavity is subjected to high circulating power. Each ring-down measurement is recorded after the laser has been locked to the cavity for a variable amount time. This is done by monitoring the cavity transmission on PD1 while scanning the laser frequency across the TEM$_{00}$ resonance peak. At resonance, the oscilloscope sends a trigger signal to the AOM driver that turns off the laser light for $500~\mu$s. The turn off time is $12.8\pm 0.2$~ns. 
This procedure is used for recording both the $455$~nm and $671$~nm cavity decay. Regarding the data analysis, an exponential decay function of the form of Eq.~\eqref{eq:transmission} is fitted to the ring-down data. The cavity lifetime $\tau$ is extracted from the fit, and the finesse is quantified via Eq.~\eqref{eq:finesse}. The results are reported in detail in the section \ref{results}.

\subsection{Physical Models of Finesse Degradation}

To characterize the temporal evolution of the cavity finesse, we compare two models for the increase in optical loss: an exponential saturating model motivated by oxygen depletion~\cite{Gangloff:15, Rudelis:23} and an Avrami model~\cite{Shirzad2023} motivated by nucleation-and-growth kinetics. These models provide two alternative descriptions of the observed dynamics, without assuming that fitting either functional form uniquely identifies the underlying microscopic degradation mechanism.

We write the total round-trip power loss as
\begin{equation}
    \ell_{\mathrm{rt}}(t) = \ell_{\mathrm{rt},0} + \Delta\ell_{\mathrm{rt}}(t),
\end{equation}
where $\ell_{\mathrm{rt},0}=\ell_{\mathrm{rt}}(0)$ is the total loss at the beginning of the measurement, including mirror transmission, absorption, and scattering, and $\Delta\ell_{\mathrm{rt}}(t) =\ell_{\mathrm{rt}}(t)-\ell_{\mathrm{rt},0}$ is the time-dependent change in round-trip loss. Both quantities are dimensionless power fractions. In particular, $\Delta\ell_{\mathrm{rt}}(0)=0$.

Using the high-finesse relation $\mathcal{F}(t)\simeq2\pi/\ell_{\mathrm{rt}}(t)$~\cite{yariv1989}, and defining $\mathcal{F}_0=\mathcal{F}(0)$, we obtain the following evolution of the finesse normalized to its initial value $\mathcal{F}_0$
\begin{equation}
    \frac{\mathcal{F}(t)}{\mathcal{F}_0}
    \simeq
    \left[
        1+\frac{\Delta\ell_{\mathrm{rt}}(t)}
        {\ell_{\mathrm{rt},0}}
    \right]^{-1}.
    \label{eq:normalized-finesse-loss}
\end{equation}

Gangloff et al.~\cite{Gangloff:15} modeled vacuum-induced degradation of tantalum-oxide-terminated mirrors through the formation of oxygen vacancies that increase optical absorption. At fixed temperature, their model assumes that the vacancy concentration approaches a limiting value via first-order kinetics, with the resulting optical loss directly proportional to this concentration. 
By expressing this model in terms of round-trip loss, one gets~\cite{Gangloff:15}
\begin{equation}
    \Delta\ell_{\mathrm{rt}}(t) = \Delta\ell_{\mathrm{O}_2,\infty} \left[ 1-\exp\left(-\frac{t}{\tau_{\mathrm{O}_2}}\right) \right],
    \label{eq:oxygen-loss}
\end{equation}
where the increment of losses $\Delta\ell_{\mathrm{O}_2,\infty} =\lim_{t\rightarrow\infty}\Delta\ell_{\mathrm{rt}}(t)>0$ is the additional round-trip loss in the limit of long exposure to vacuum as predicted by this model. The parameter $\tau_{\mathrm{O}_2}>0$ is the characteristic depletion timescale. In their work, Gangloff et al. describe their model in terms of excess loss per mirror. For two mirrors undergoing the same loss increase, the corresponding increment of round-trip loss is twice the single-mirror increment, assuming that the mirror transmission and losses remain constant. By defining the dimensionless relative loss amplitude as
\begin{equation}
    a_{\mathrm{O}_2} \equiv \frac{\Delta\ell_{\mathrm{O}_2,\infty}}
    {\ell_{\mathrm{rt},0}},
    \label{eq:oxygen-amplitude}
\end{equation}
we obtain the following time evolution of the finesse due to oxygen vacancies
\begin{equation}
    \frac{\mathcal{F}_{\mathrm{O}_2}(t)}
         {\mathcal{F}_0} \simeq \frac{1}{1+a_{\mathrm{O}_2}
        \left[ 1-\exp\left(-t/\tau_{\mathrm{O}_2}\right) \right]}.
    \label{eq:oxygen-finesse}
\end{equation}

The parameters $a_{\mathrm{O}_2}$ and $\tau_{\mathrm{O}_2}$ describe the magnitude and timescale of the loss increase, respectively. The predicted steady-state finesse is $\mathcal{F}_{\infty} \simeq\mathcal{F}_0/(1+a_{\mathrm{O}_2})$.

As an alternative, we can describe the finesse degradation using an Avrami-type model, motivated by nucleation-and-growth kinetics~\cite{Shirzad2023}:
\begin{equation}
    \frac{\mathcal{F}_{\mathrm{A}}(t)}{\mathcal{F}_0} = \frac{1}{1+a_{\mathrm{A}} \left[ 1-\exp\left[-(t/\tau_{\mathrm{A}})^n\right] \right]}.
    \label{eq:avrami-finesse}
\end{equation}

Here, $\tau_{\mathrm{A}}>0$ is the characteristic timescale, and $n$ is the Avrami exponent, which controls the temporal shape of the loss increase. The dimensionless amplitude $a_{\mathrm{A}}=\Delta\ell_{\mathrm{A},\infty}/ \ell_{\mathrm{rt},0}>0$ is the ratio of the additional round-trip losses in the long-time limit to the initial total losses. The equation follows from Eq.~\eqref{eq:normalized-finesse-loss} by assuming that the additional loss term $\Delta\ell_{\mathrm{rt}}(t)$ evolves according to the Avrami equation. In our data analysis, we fix the Avrami exponent at $n=3$, a value consistent with continuous nucleation and two-dimensional, interface-controlled growth of defects on a mirror surface~\cite{Shirzad2023}. This is because, in this model, new defect regions appear at a constant rate with their dimensions increasing linearly with time. The area of an individual region, therefore, grows quadratically over time. To this contribution one has to add the continuous formation of new defect regions, giving an Avrami exponent of three. This provides a possible physical picture for the formation and lateral growth of loss-inducing regions on a mirror surface. 

In our experiments, we use both the oxygen depletion (Eq.~\eqref{eq:oxygen-finesse}) and the Avrami/Nucleation (Eq.\eqref{eq:avrami-finesse}) models to fit the data of the finesse degradation.


\section{Results}\label{results}

We performed a series of experiments to investigate how prolonged resonant illumination affects the cavity finesse at 455 nm and 671 nm. These measurements address both the wavelength dependence of degradation in vacuum and the recovery of the finesse under illumination in air. We considered three experimental conditions:

\begin{itemize}
    \item Resonant illumination at 455 nm in vacuum, to characterize degradation at 455 nm and its possible impact on the finesse at 671 nm.
    \item Resonant illumination at 671 nm in vacuum, to assess the stability of the finesse at both wavelengths.
    \item Resonant illumination at 455 nm in air, to investigate the recovery of the blue finesse while monitoring any changes in the finesse at 671 nm.
\end{itemize}



\subsection{Finesse degradation at $455$~nm}

\begin{figure}[ht]
    \centering
    \includegraphics[width=0.9\linewidth]{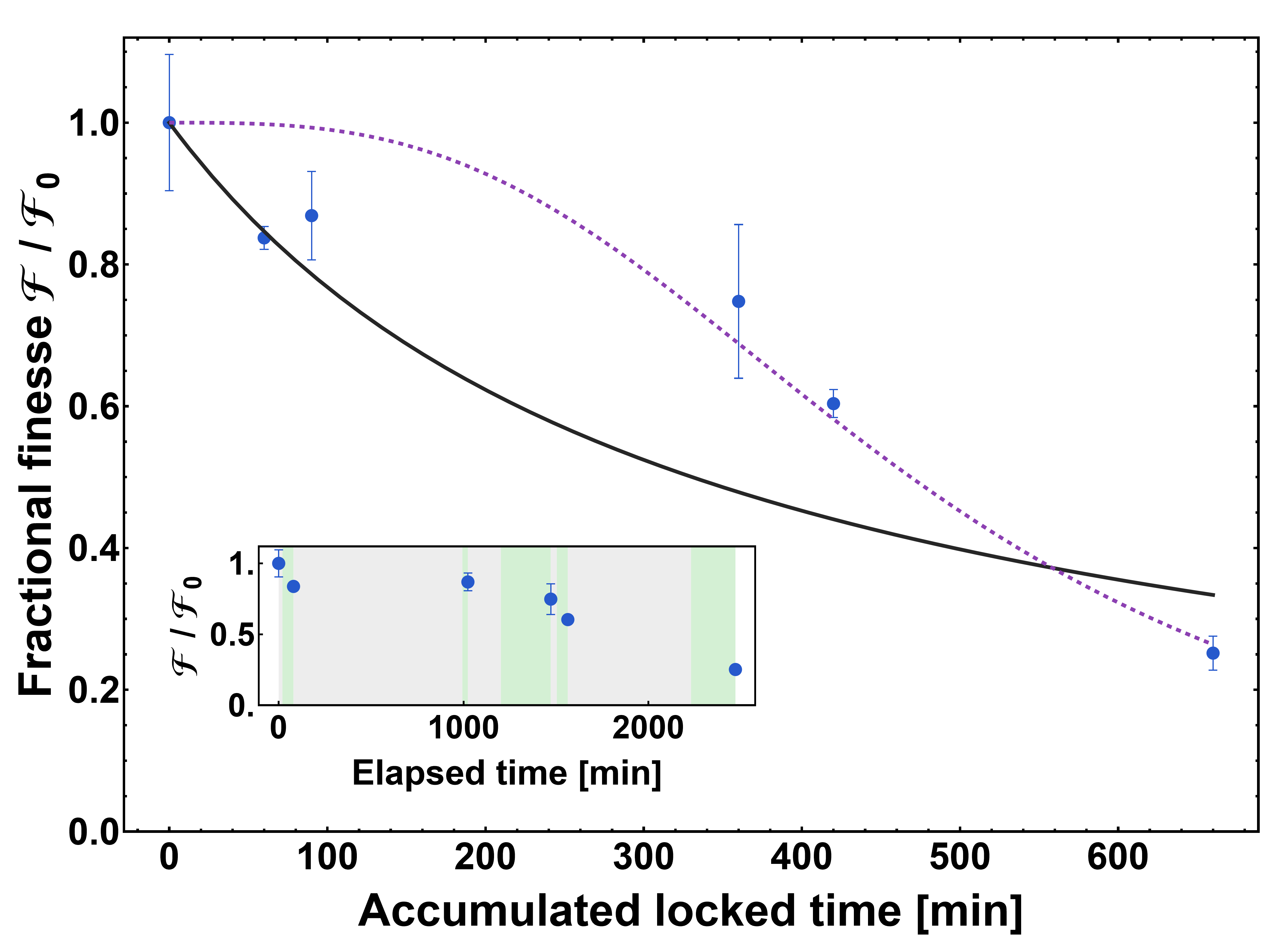}
    \caption{Cavity finesse degradation at 455~nm. Evolution of the cavity finesse while locked at 455~nm under vacuum, with an output power of $P_{\text{out},455}=142.8~\mu$W. The finesse is referenced to the initial measured value $\mathcal{F}_0=(7.7\pm 0.7)\times 10^3$. Each data point corresponds to an average of 5 ring-down measurements. The error bars indicate one standard deviation error of the mean of the respective averaged measurements. Each ring-down measurement at $455$~nm was fitted from the turn-off time $t_0$ following Eq.\eqref{eq:transmission}. The progressive decrease in finesse is fitted to the nucleation/Avrami model in Eq.~\eqref{eq:avrami-finesse} (purple curve) and to the oxygen-depletion model in Eq.~\eqref{eq:oxygen-finesse} (orange curve) are also plotted The inset shows the same data  as a function of elapsed time, where the time intervals in which the laser was locked to the cavity (green) or not locked to the cavity (gray) are highlighted. The cavity remained in vacuum at all times. 
    }
    \label{fig:455_deg_1}
\end{figure}

First, we tested the cavity finesse degradation at $455$~nm by locking this laser to the optical cavity for an extended period of time. The input power is set to $P_{\text{in},455}=1.83$~mW, whereas the output power measured in transmission is on average $P_{\text{out},455}=142.8~\mu$W. Initial finesse is $\mathcal{F}_0=(7.7\pm 0.7)\times 10^3$.
We observed a 75\% decrease in finesse, which degraded down to a final value of $\mathcal{F}_f=(1.9\pm 0.2)\times 10^3$
after $660$~mins ($11$~hours) of illumination, as shown in Fig.~\ref{fig:455_deg_1}. The finesse data is normalized with respect to the maximum finesse value, $\mathcal{F}_{\text{max}}$, that in this case, is $\mathcal{F}_0$. To analyze the finesse evolution we applied two degradation models, a nucleation/Avrami model, and a oxygen depletion model, both introduced in the previous section. As Fig.~\ref{fig:455_deg_1} shows, the nucleation/Avrami model reproduces the overall degradation trend, whereas the oxygen-depletion model fails to capture the observed decay. Moreover, during the time where the laser is not locked to the cavity (no circulating power), we observe no decrease in the finesse, see inset in figure. Our findings are suggestive of hydrocarbon contamination as cause for cavity degradation.

\subsection{Effect of the finesse degradation at $455$~nm on the finesse at $671$~nm}

\begin{figure}[ht]
    \centering
    \includegraphics[width=1\linewidth]{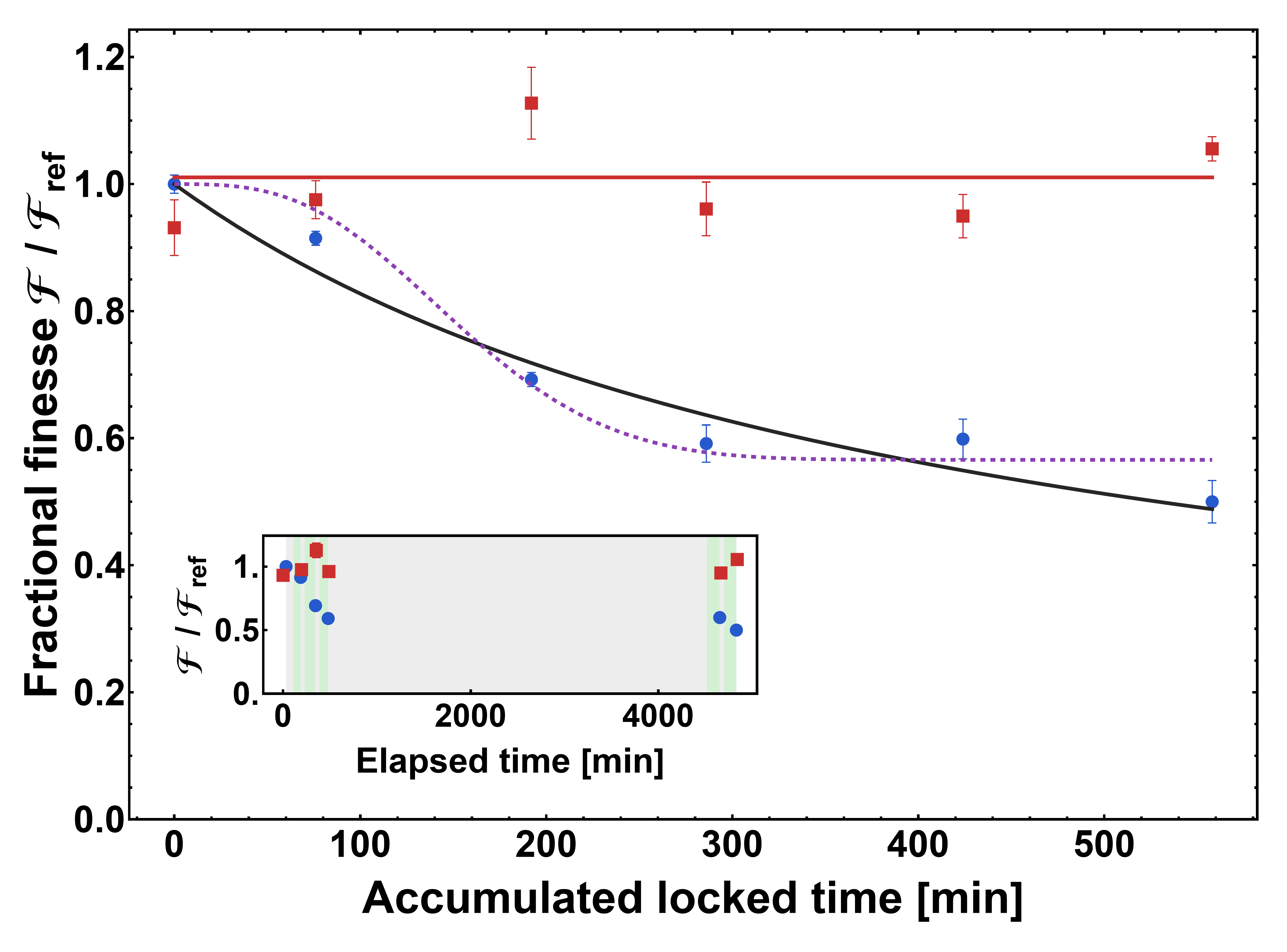}
    \caption{Finesse evolution at $455$~nm and $671$nm while locking at $455$~nm under vacuum. $455$~nm output power is $P_{\text{out},455}=38.3~\mu$W. Finesse values at $455$~nm (blue data) were normalized with respect to the initial finesse $\mathcal{F}_{ref}=\mathcal{F}_0=(4.82\pm 0.07)\times 10^3$, while finesse values at $671$~nm (red data) where normalized with respect to average value $\mathcal{F}_{ref}=\bar{\mathcal{F}}_{671}=(1.6\pm 0.1)\times 10^4$. The first 3 data points represent the average of 5 measurements, while the last three correspond to an average of 10 measurements. The error bars indicate one standard deviation. While the finesse at 455~nm progressively decreases, the finesse at $671$~nm remains unchanged throughout the measurement. The finesse depletion at $455$~nm is more consistent with the nucleation model (purple curve) than with the oxygen-depletion model (orange curve). The inset shows the same data as a function of elapsed time, where the time intervals in which the laser was locked to the cavity (green) or not locked to the cavity (gray) are highlighted. The cavity remained in vacuum at all times.
    }
    \label{fig:455/671_deg_455}
\end{figure}

After partial recovery of the cavity finesse with $455$~nm light illumination in air, see section~\ref{recovery}, we run a second experiment in order to test the effect of the degradation caused by $455$~nm illumination on the finesse measured at $671$~nm. We lock again the $455$~nm laser on the cavity and observe the evolution of the finesse at both wavelengths. The average input power is $P_{\text{in},455}=1.37$~mW and $P_{\text{in},671}=9.18$~mW, with corresponding  average output power at $455$~nm of $P_{\text{out},455}=38.3~\mu$W. The initial $455$~nm finesse value is $\mathcal{F}_0=(4.82\pm 0.07)\times 10^3$, and this value
is used to normalize the $455$~nm finesse data. The evolution of the fractional change in finesse at both wavelengths is shown in Fig.~\ref{fig:455/671_deg_455}. As in the previous run, we observe a degradation of the finesse at $455$nm to the value $\mathcal{F}_f=(2.4\pm 0.2)\times 10^3$ 
over $558$~mins ($9.3$~hours) of locking time. The finesse at $671$~nm, however, does not show degradation as the finesse data fluctuate around a mean value of  $\bar{\mathcal{F}}_{671}=(1.6\pm 0.1)\times 10^4$, used to normalize the data. The data of $455$~nm degradation are fitted with the nucleation and oxygen depletion models. In this set of data both models qualitatively agree with the trend, nucleation model still being the one with lowest variance and therefore best fitting. 
After data acquisition, we found that ring-down curves were convolved with the instrumentation time-response due to an internal low-pass filter in the oscilloscope.
To account for this convolution, we fitted the ring-down data over two normalized amplitude windows centered around $50$\% of the decay: from $0.66$ to $0.34$ and from $0.815$ to $0.185$. The finesse reported in Fig.~\ref{fig:455/671_deg_455} is the mean of the two estimates, and its uncertainty is the mean of their uncertainties.

\subsection{Effect of continuous illumination at $671$~nm on the finesse at $455$~nm}
\begin{figure}[ht]
    \centering
    \includegraphics[width=1\linewidth]{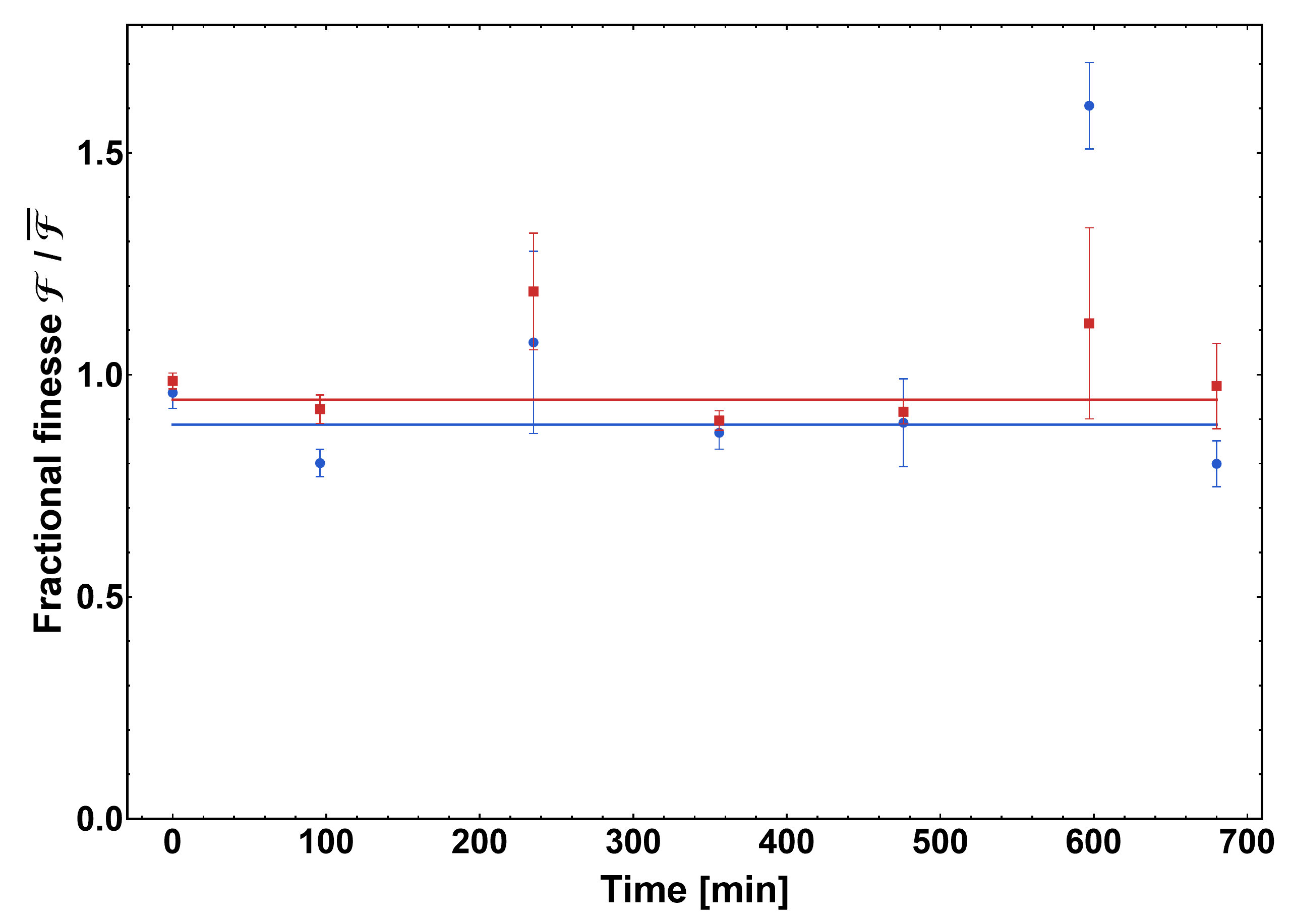}
    \caption{Cavity finesse stability under continuous illumination at $671$~nm. Evolution of the cavity finesse at $455$~nm (blue points) and $671$~nm (red squares) when the cavity is locked at $671$~nm, with an output power of $P_{\text{out},671}=700~\mu$W from the cavity when locked. Finesse values at both wavelengths are normalized with respect to the mean finesse values $\mathcal{F}_{ref}=\bar{\mathcal{F}}_{455}=(2.5\pm 0.7)\times 10^3$ and $\mathcal{F}_{ref}=\bar{\mathcal{F}}_{671}=(1.7\pm 0.2)\times 10^4$, respectively. The blue and red lines indicate the weighted average finesse at $455$~nm and $671$~nm respectively. No measurable degradation is observed at either wavelength despite the high $671$~nm circulating power. Each data point represents the average of 10 measurements and the error bars indicate one standard deviation of the averaged ring-down measurements.}
    \label{fig:455/671_deg_671}
\end{figure}

After observing the degradation with high $455$~nm circulating intensity, we tested the cavity degradation effect when locking the $671$~nm laser to the cavity. We monitor the evolution of the finesse at both wavelengths. This measurement was done with two different conditions. In the first one, we set the input powers at $P_{\text{in},455}=1.12$~mW and $P_{\text{in},671}=8.77$~mW. The average power out of the cavity measured at $671$~nm is $P_{\text{out},671}=700~\mu$W. 
In this case the experiment was done without previously recovering the finesse at $455$~nm. During this measurement, the $671$~nm laser was locked to the cavity for a total time of $680$~mins ($11.3$~hours). The results are shown in Fig.~\ref{fig:455/671_deg_671}. Despite the high circulating power at $671$~nm, the data show no finesse degradation, since the finesse values at both wavelength fluctuate around a certain mean value, measured to be $\bar{\mathcal{F}}_{455}=(2.5\pm 0.7)\times 10^3$ and $\bar{\mathcal{F}}_{671}=(1.7\pm 0.2)\times 10^4$. 

On a second set of measurements, we repeated the same experiment after a partial recovery of the cavity finesse at $455$~nm. In this case, the input powers of both beams were on average $P_{\text{in},455}=0.976$~mW and $P_{\text{in},671}=6.15$~mW. The average output power at $671$~nm was measured to be $P_{\text{out},671}=503~\mu$W.
We observed once more that the finesse at both wavelengths fluctuates around an average value, found to be $\bar{\mathcal{F}}_{455}=(3.4\pm 0.9)\times 10^3$ and $\bar{\mathcal{F}}_{671}=(1.8\pm 0.3)\times 10^4$, even when the cavity finesse has been recovered. This time the laser was locked to the cavity for a total time of $812$~mins ($13.5$~hours). These two measurements show that no coating degradation is caused by $671$~nm light running into the cavity and, moreover, they confirm once more that there is no evidence of a degradation purely caused by having the cavity under vacuum.


\subsection{Finesse recovery}\label{recovery}

\begin{figure}[ht]
    \centering
    \includegraphics[width=1\linewidth]{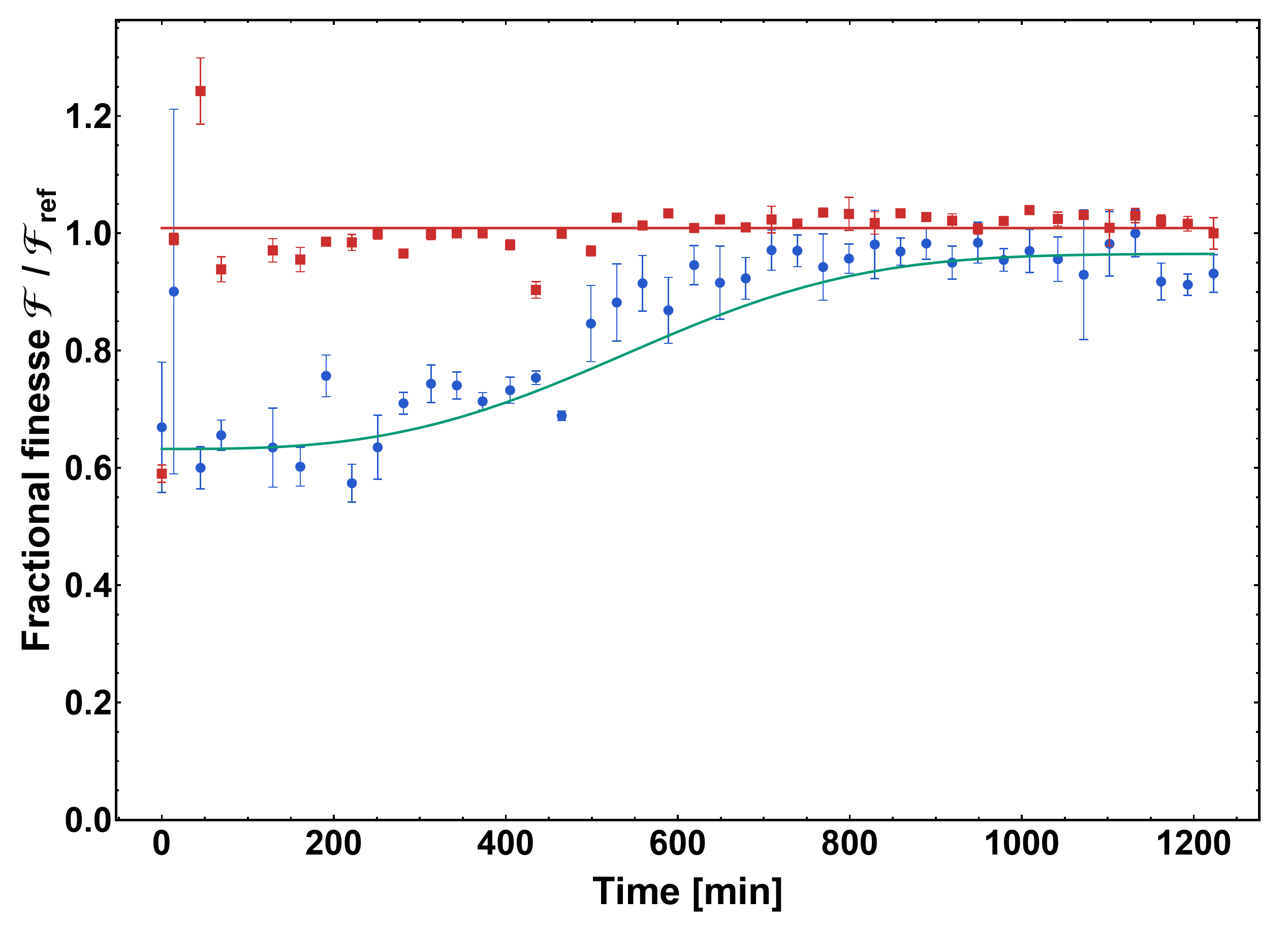}
    \caption{Cavity finesse recovery in air with $455$~nm laser locked to the cavity. Blue points correspond to the  cavity finesse at $455$~nm, while red squares represent the finesse at $671$~nm, both measured with cavity ring-down. Each data point represents the average of 10 measurements and the error bars indicate one standard deviation. We observe a recovery of the cavity finesse at $455$~nm when the laser is locked to the cavity in air, with an average output power of $P_{\text{out},455}=18.07~\mu$W. Finesse values at $455$~nm are normalized with the maximum measured finesse, in this case $\mathcal{F}_{ref}=\mathcal{F}_{max}=(3.1\pm 0.1)\times 10^3$. Instead, $671$~nm finesse values are normalized with respect to the mean finesse $\mathcal{F}_{ref}=\bar{\mathcal{F}}_{671}=(1.3\pm 0.1)\times 10^4$ (red line). The finesse at $671$~nm remains unchanged throughout the measurement. The recovery (green line) are fitted with the phenomenological saturation growth given in Eq.~\eqref{eq:rec_model}. 
    }
    \label{fig:455/671_recovery}
\end{figure}

Following the work done by Ref.~\cite{Gangloff:15}, the cavity finesse was partially recovered by locking the $455$~nm laser to the cavity in air. During the procedure we monitor both the $455$~nm finesse for checking the recovery and that at $671$~nm with ring down measurement in order to monitor any potential effect of the recovery on the $671$~nm finesse. The average input power of both lasers used during the oxygen recovery is $P_{\text{in},455}=1.29$~mW and $P_{\text{in},671}=8.32$~mW. The average output power at $455$~nm was measured to be $P_{\text{out},455}=18.1~\mu$W.
The results of the recovery are shown in Fig.~\ref{fig:455/671_recovery}. The initial finesse value at $455$~nm is $\mathcal{F}_0=(2.0\pm 0.3)\times 10^3$. After the recovery which lasts a total illumination time of $1223$~mins ($20.4$~hours), it reaches a steady state final value of $\mathcal{F}_f=(2.8\pm 0.1)\times 10^3$. 
The finesse at $671$~nm, instead, fluctuates around an average value of $\bar{\mathcal{F}}_{671}=(1.3\pm 0.1)\times 10^4$ during the whole recovery. In Fig.~\ref{fig:455/671_recovery} the finesse values at $455$~nm are normalized with respect to the maximum finesse value measured during recovery, $\mathcal{F}_{max}=(3.1\pm 0.1)\times 10^3$. The finesse values at $671$~nm, instead, are normalized with respect to the average value obtained from the whole set of data. The data of the $455$~nm recovery are found to fit a model of the form
\begin{equation}
    \mathcal{F}(t) = \mathcal{F}_{\infty}-(\mathcal{F}_{\infty}-\mathcal{F}_0)\text{exp}\left[-\left(\frac{t}{\tau_{\text{rec}}}\right)^3\right],
    \label{eq:rec_model}
\end{equation}
where $\mathcal{F}_{\infty}$ is the asymptotic finesse associated to the recovery process, $\mathcal{F}_0$ is the initial finesse and $\tau_{\text{rec}}$ is the characteristic recovery time. The recovery is described phenomenologically by a cubic-exponent saturation function. An exponent of three is compatible with continuous nucleation and two-dimensional growth under classical Avrami assumptions, however, neither these spatial kinetics nor a linear relationship between recovered fraction and finesse is independently established in this work.

\section{Conclusions}
In this work, we investigated the degradation and recovery of a bichromatic Fabry-Perot cavity resonant at both $455$~nm and $671$~nm. Under vacuum, with the cavity locked at $455$~nm and operating at moderate input optical power, the first degradation run showed a decrease in finesse of about 75\%, from $\mathcal{F}_0=(7.7\pm 0.7)\times 10^3$ to $\mathcal{F}_f=(1.9\pm 0.2)\times 10^3$, after approximately 11 hours of accumulated illumination. Comparison of oxygen-depletion and Avrami-type models showed that the observed degradation dynamics were better described by the Avrami model, consistent with nucleation-and-growth kinetics. Moreover, no measurable finesse degradation was observed during intervals without circulating optical power, suggesting that the vacuum environment alone is insufficient to account for the observed degradation. The final finesse in steady state suggest that that the cavity can be operated safely with an intensity of 
$I_{455}\sim5$~kW$/$cm$^2$ 
on the cavity mirrors.

Despite the substantial degradation observed at $455$~nm, the finesse at $671$~nm remained unchanged within experimental uncertainty. This suggests that the processes responsible for the increased optical losses at $455$~nm  have little or no measurable effect on the cavity finesse at $671$~nm. Furthermore, when high $671$~nm power was circulating in the cavity under vacuum, no measurable degradation was observed at either wavelength. This finding confirmed once more that the degradation is caused by blue radiation circulating in the cavity under vacuum, suggestive of degradation due to hydrocarbon contamination. 
Finally, a partial recovery of the finesse at $455$~nm was observed with the cavity locked at $455$~nm in air.

These results demonstrate that high circulating powers in the red ($671$~nm) can be safely used for applications such as optical trapping and cavity locking, without measurable degradation of the cavity finesse under the conditions investigated. In contrast, blue light ($455$~nm) applications should aim to limit the circulating photon number, for example by using short, low-power pulses or restricting blue light illumination to applications requiring enhanced coupling to the vacuum electromagnetic field. Minimizing potential sources of contamination inside the vacuum chamber~\cite{Schmitz2019} or performing annealing of the cavity mirrors~\cite{Brandstatter2013} may also help mitigate long-term degradation under continuous blue laser light illumination. These findings are very relevant for the design of experiments with optical cavities that need to be used both for trapping and for cavity coupling, or for CQED with atomic mixtures.

\section{Acknowledgments}
We thank M. Keller for helpful discussions. This work has been supported by the project 23FUN03 HIOC, which has received funding from the European Partnership on Metrology, cofinanced from the European Union’s Horizon Europe Research and Innovation Programme and by the Participating States. This work was supported by the Cascading Grant Spoke 3 under the PNRR MUR project PE0000023-NQSTI, funded by the European Union – Next Generation EU.

The raw data used in this manuscript are publicly available in Ref. \cite{https://doi.org/10.5281/zenodo.23798276}.

\nocite{*}
\bibliography{biblio}

\end{document}